\documentclass[%
 aip,
 amsmath,amssymb,
 reprint,%
]{revtex4-1}

\usepackage{graphicx}
\usepackage{dcolumn}
\usepackage{bm}

\usepackage[utf8]{inputenc}
\usepackage[T1]{fontenc}
\usepackage{mathptmx}
\usepackage{etoolbox}

\usepackage[
CJKbookmarks=false,
colorlinks,
breaklinks=true,
linkcolor=blue,
anchorcolor=blue,
citecolor=blue,
urlcolor=blue,
]{hyperref}
\usepackage{textcomp}
\usepackage{subfigure}

\makeatletter
\def\@email#1#2{%
 \endgroup
 \patchcmd{\titleblock@produce}
  {\frontmatter@RRAPformat}
  {\frontmatter@RRAPformat{\produce@RRAP{*#1\href{mailto:#2}{#2}}}\frontmatter@RRAPformat}
  {}{}
}%
\makeatother

\begin{document}

\preprint{AIP/123-QED}

\title[]{Mitigating Measurement Crosstalk via Pulse Shaping}

\author{Yang Gao}
\thanks{ These authors contributed equally to this work.} 
\affiliation{Beijing Key Laboratory of Fault-Tolerant Quantum Computing, Beijing Academy of Quantum Information Sciences, Beijing 100193, China}
\affiliation{Beiing National Laboratory for Condensed Matter Physics, institute of Physics, Chinese Academy of sciences, Beiing 100190, China}
\affiliation{University of Chinese Academy of Sciences, Beiiing 100049, China}

\author{Feiyu Li}
\thanks{ These authors contributed equally to this work.} 
\affiliation{Beijing Key Laboratory of Fault-Tolerant Quantum Computing, Beijing Academy of Quantum Information Sciences, Beijing 100193, China}
\affiliation{Beiing National Laboratory for Condensed Matter Physics, institute of Physics, Chinese Academy of sciences, Beiing 100190, China}
\affiliation{University of Chinese Academy of Sciences, Beiiing 100049, China}

\author{Yang Liu}
\affiliation{Beijing Key Laboratory of Fault-Tolerant Quantum Computing, Beijing Academy of Quantum Information Sciences, Beijing 100193, China}

\author{Zhen Yang}
\affiliation{Beijing Key Laboratory of Fault-Tolerant Quantum Computing, Beijing Academy of Quantum Information Sciences, Beijing 100193, China}

\author{Jiayu Ding}
\affiliation{Beijing Key Laboratory of Fault-Tolerant Quantum Computing, Beijing Academy of Quantum Information Sciences, Beijing 100193, China}

\author{Wuerkaixi Nuerbolati}
\affiliation{Beijing Key Laboratory of Fault-Tolerant Quantum Computing, Beijing Academy of Quantum Information Sciences, Beijing 100193, China}

\author{Ruixia Wang}
\affiliation{Beijing Key Laboratory of Fault-Tolerant Quantum Computing, Beijing Academy of Quantum Information Sciences, Beijing 100193, China}

\author{Tang Su}
\affiliation{Beijing Key Laboratory of Fault-Tolerant Quantum Computing, Beijing Academy of Quantum Information Sciences, Beijing 100193, China}

\author{Yanjun Ma}
\affiliation{Beijing Key Laboratory of Fault-Tolerant Quantum Computing, Beijing Academy of Quantum Information Sciences, Beijing 100193, China}

\author{Yirong Jin}
\affiliation{Beijing Key Laboratory of Fault-Tolerant Quantum Computing, Beijing Academy of Quantum Information Sciences, Beijing 100193, China}

\author{Haifeng Yu}
\affiliation{Beijing Key Laboratory of Fault-Tolerant Quantum Computing, Beijing Academy of Quantum Information Sciences, Beijing 100193, China}
\affiliation{Hefei National Laboratory, Hefei 230088, China}

\author{He Wang\textsuperscript{*}}
\email{wanghe@baqis.ac.cn}
\affiliation{Beijing Key Laboratory of Fault-Tolerant Quantum Computing, Beijing Academy of Quantum Information Sciences, Beijing 100193, China}

\author{Fei Yan}
\affiliation{Beijing Key Laboratory of Fault-Tolerant Quantum Computing, Beijing Academy of Quantum Information Sciences, Beijing 100193, China}

\date{\today}

\begin{abstract}
Quantum error correction protocols require rapid and repeated qubit measurements.  While multiplexed readout in superconducting quantum systems improves efficiency, fast probe pulses introduce spectral broadening, leading to signal leakage into neighboring readout resonators. This crosstalk results in qubit dephasing and degraded readout fidelity. Here, we introduce a pulse shaping technique inspired by the derivative removal by adiabatic gate (DRAG) protocol to suppress measurement crosstalk during fast readout. By engineering a spectral notch at neighboring resonator frequencies, the method effectively mitigates spurious signal interference. Our approach integrates seamlessly with existing readout architectures, enabling fast, low-crosstalk multiplexed measurements without additional hardware overhead---a critical advancement for scalable quantum computing.
\end{abstract}

\maketitle

Readout of superconducting qubits is a key element in superconducting quantum processors \cite{divincenzo2000physical,blais2021circuit}. The predominant dispersive readout approach \cite{wallraff2004strong}, where a qubit is coupled to a resonator whose frequency is probed with coherent microwave fields, has achieved high performance in both fidelity and speed \cite{jerger2012frequency,jeffrey2014Fast,Walter2017Rapid,heinsoo2018rapid,sunada2022fast,chen2023transmon,swiadek2024enhancing,spring2025fast}. In scalable architectures, multiple resonators are often coupled to a common feedline, enabling frequency-multiplexed readout \cite{schmitt2014multiplexed,heinsoo2018rapid,spring2025fast}, which reduces wiring complexity and enhances scalability. Moreover, the ability to perform rapid readout is particularly critical for applications such as quantum error correction, where idling errors in data qubits---arising while waiting for ancilla qubits to be measured---represent a major limitation \cite{google2023suppressing,google2025quantum}. State-of-the-art superconducting systems now achieve qubit readout within tens of nanoseconds \cite{spring2025fast}.

Sharing a common feedline can introduce crosstalk between resonators: a microwave pulse intended for one resonator can leak into others, unintentionally inducing additional readout errors, qubit frequency shifts, and qubit dephasing \cite{gambetta2006qubit,heinsoo2018rapid,Sunada2024Photon}. This crosstalk issue becomes particularly significant in fast multiplexed readout, where resonator linewidths $\kappa$ are large and probe pulses are short \cite{swiadek2024enhancing,heinsoo2018rapid,spring2025fast}. Larger resonator linewidths increase spectral overlap between adjacent resonators, while shorter pulses broaden the spectral width of probe signals. In multiplexed systems, these two effects jointly contribute to readout crosstalk. 

\begin{figure}[tb]
    \centering
    \includegraphics[width=1\linewidth]{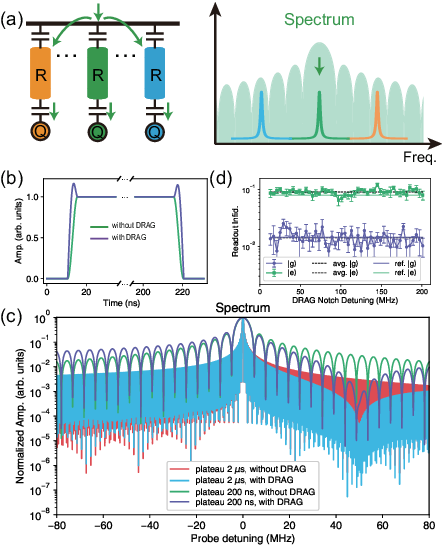}
    \subfigure{\label{fig:Figure1Concept:a}}
    \subfigure{\label{fig:Figure1Concept:b}}
    \subfigure{\label{fig:Figure1Concept:c}}
    \subfigure{\label{fig:Figure1Concept:d}}
    \caption{
    \textbf{Removing probe signal leakage using DRAG.} 
    (a) Schematic of multiplexed readout: multiple resonators share a common feedline. The spectrum of the probe pulse overlaps with the resonance of untargeted readout resonators, leading to additional crosstalk errors and qubit dephasing. 
    (b) Time-domain envelopes of readout pulses with cosine-shaped rise and fall edges and a 200-ns flat plateau, shown with and without DRAG. 
    (c) Effect of DRAG in suppressing the 50-MHz spectral component for readout pulses with plateau durations of 200~ns and 2~\textmu s. For the 2 \textmu s case, the spectral component at 50 MHz is suppressed by over an order of magnitude when the DRAG pulse is applied.
    (d) Dependence of readout infidelity on the DRAG notch frequency. The readout pulse plateau is 2~\textmu s. The solid reference line indicates the readout infidelity without applying DRAG.
    }
    \label{fig:concept}
\end{figure}

Readout crosstalk can be mitigated by increasing the frequency spacing between resonators; however, larger spacing imposes stricter bandwidth requirements on components across the readout chain, such as Josephson parametric amplifiers (JPAs) \cite{mutus2013design,mutus2014strong}. Replacing JPAs with broadband traveling-wave parametric amplifiers \cite{wang2025highefficiencylowlossfloquetmodetraveling,Sandbo2025Josephson} could largely alleviate this constraint. Nevertheless, the substantial increase in qubit count required for fault-tolerant application-scale quantum computers \cite{eisert2025mind} demands dense integration of readout resonators, thereby posing the challenge of improving the multiplexing capability within a finite system bandwidth. Another approach is to equip each readout resonator with a dedicated Purcell filter \cite{heinsoo2018rapid}, which suppresses off-resonant driving and thereby reduces spectral overlap between resonators. Nevertheless, spurious errors induced by short readout pulses that spectrally extend to the resonance of untargeted resonators [Fig.~\ref{fig:Figure1Concept:a}] cannot be eliminated by this method. To address this limitation, we introduce a pulse-shaping technique: adding a derivative component to the probe pulse [Fig.~\ref{fig:Figure1Concept:b}] to suppress targeted frequency components [Fig.~\ref{fig:Figure1Concept:c}]. This derivative removal by adiabatic gate (DRAG) method \cite{Motzoi2009Simple}, widely used in qubit control to suppress leakage \cite{chen2016measuring,hyyppa2024reducing}, is here adapted to suppress unwanted spectral ingredients at neighboring resonator frequencies, thereby reducing errors arising from the broad bandwidth of short pulses. Our method is hardware-friendly at the system level, as it introduces no additional physical components, although it does require a higher output amplitude from the waveform generator---a requirement that most current devices can readily meet. The method is well-suited for architectures employing individual Purcell filters, in which spectral overlap between readout resonators is already small and readout crosstalk is instead dominated by the broad spectral width of the probe pulses.

Our method is based on the DRAG technique. Given a probe pulse envelope $W(t)$, we add the time derivative
$\dot{W}(t)$ to the quadrature component:
\begin{equation}
W_{\rm drag}(t) = W(t) + i \frac{\dot{W}(t)}{\eta},
\end{equation}
where $\eta$ is the detuning between the resonator center frequency and the engineered notch frequency, as illustrated in Fig.~\ref{fig:Figure1Concept:b}. Fourier analysis shows that the component at $\eta = 50$ MHz is suppressed, as shown in Fig.~\ref{fig:Figure1Concept:c}. In multiplexed readout systems, we can remove spurious spectral ingredients at neighboring resonator frequencies, thereby mitigating measurement crosstalk.

In this work, the experiments are performed on a qubit-resonator system, with the resonator frequency being $\omega_r/2\pi = 6.901$~GHz and the qubit frequency fixed at $\omega_q/2\pi = 4.392$~GHz. The resonator decay rate is $\kappa/2\pi = 2.2$~MHz, and the dispersive shift is $2\chi/2\pi = 2.1$~MHz. The qubit energy relaxation time $T_1$ varied between 92-101~\textmu s, while the transverse relaxation time $T_2$ ranged from 15-18~\textmu s, based on repeated measurements conducted on separate days using standard $T_1$ and Ramsey protocols.

Firstly, we evaluated the effect of DRAG on the qubit's readout fidelity by comparing the cases with and without DRAG [Fig.~\ref{fig:Figure1Concept:d}]. A plateau duration of 2 \textmu s was used to ensure sufficient interaction time for achieving a high signal-to-noise ratio (SNR). The SNR could be further improved by using optimal weight functions instead of square weight functions \cite{bultink2018general}. The solid reference line indicates the readout infidelity without applying DRAG on the probe pulse. The results demonstrate that applying DRAG introduces a 0.6\% reduction in the average readout fidelity when the notched-frequency detuning is scanned across 13-201 MHz. Furthermore, the effect of DRAG at higher fidelity levels (such as above 99\%) remains to be investigated.

\begin{figure}[tb]
    \centering
    \includegraphics[width=\linewidth]{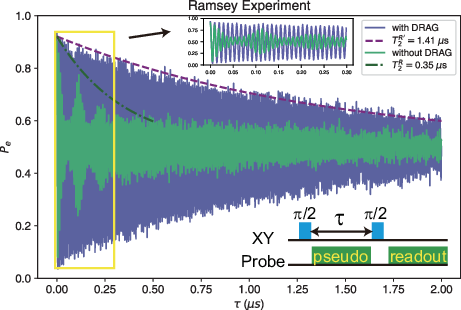}
    \caption{
    \textbf{Recovering phase coherence by DRAG.}
    Ramsey measurement results with and without DRAG applied to the pseudo-readout pulse. The pulse is detuned by $+10$~MHz from the readout resonator, with a flat plateau of duration $\tau$ and 10-ns cosine-shaped rise and fall edges. The final readout pulse plateau is 2~\textmu s. Without DRAG, a clear frequency beating pattern appears (inset), whereas DRAG suppresses the beating and extends the qubit coherence time.
    }
    \label{fig:ramsey}
\end{figure}

\begin{figure*}[tb]
    \centering
    \includegraphics[width=0.9\linewidth]{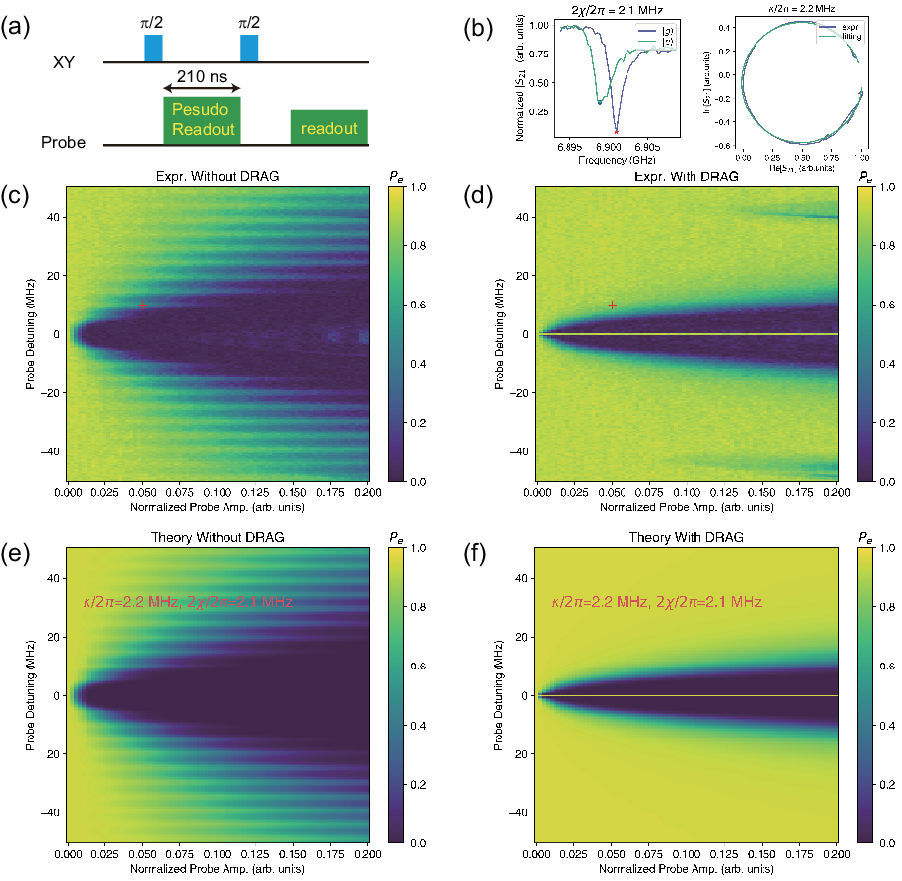}
    \subfigure{\label{fig:ProbeInducedDephasing:a}}
    \subfigure{\label{fig:ProbeInducedDephasing:b}}
    \subfigure{\label{fig:ProbeInducedDephasing:c}}
    \subfigure{\label{fig:ProbeInducedDephasing:d}}
    \subfigure{\label{fig:ProbeInducedDephasing:e}}
    \subfigure{\label{fig:ProbeInducedDephasing:f}}
    \caption{
    \textbf{Probe frequency and amplitude dependence of readout-induced qubit dephasing.}
    (a) Pulse sequence used to measure readout induced dephasing. To avoid beating in the Ramsey experiment, the pseudo-readout pulse length is fixed. The pulse envelope has a cosine shape with a flat plateau, 10-ns total rise and fall times, and a 200-ns plateau. The phase of the second $\pi/2$ gate is scanned, and the resulting oscillation is fitted to extract the amplitude and phase. The final readout pulse plateau is 2~\textmu s.
    (b) Experimentally extracted parameters: $\kappa/2\pi = 2.2$~MHz, $2\chi/2\pi = 2.1$~MHz. 
    (c-d) Experimental results of scanning the pseudo-readout pulse amplitude and detuning, without DRAG (c) and with DRAG (d). The red cross indicates the pseudo-readout pulse parameters used in Fig. \ref{fig:ramsey}.
    (e-f) Theoretical calculations of the same scans, without DRAG (e) and with DRAG (f).}
    \label{fig:ProbeInducedDephasing}
\end{figure*}

Then, we performed a Ramsey experiment to characterize the effect of a detuned probe drive on a qubit. As shown in Fig.~\ref{fig:ramsey}, a pseudo-readout pulse was inserted between two $\pi/2$ pulses, with its plateau duration $\tau$ varied to control the effective delay. The phase $\theta$ of the second $\pi/2$ pulse was swept as a function of $\tau$ to obtain the Ramsey oscillations. The pseudo-readout pulse was detuned by $+10$ MHz from the target resonator. Without DRAG, the oscillations exhibited a pronounced beating pattern, as shown in the inset of Fig.~\ref{fig:ramsey}. By contrast, applying DRAG substantially suppressed the beating and extended the qubit's $T_2$ time to $1.41$ \textmu s (compared with 0.35 \textmu s without DRAG), as extracted from an exponential fit to the oscillation envelope. This contrast depends on the amplitude and detuning of the pseudo-readout pulse, as shown in Figs. \ref{fig:ProbeInducedDephasing:c}-\ref{fig:ProbeInducedDephasing:f}.

When the pseudo readout pulse is short, its broad spectral bandwidth overlaps with the resonator frequency, causing the resonator to respond to both resonant and off-resonant components, which gives rise to the observed beating. The application of DRAG suppresses the resonant spectral component, thereby reducing both the beating and the associated qubit dephasing.

Finally, we measured the rate of readout-induced qubit dephasing, as shown in Fig.~\ref{fig:ProbeInducedDephasing}. Such measurement-induced dephasing can also serve as an effective metric for quantifying measurement crosstalk on untargeted qubits \cite{heinsoo2018rapid}. The average dephasing rate was obtained using a Ramsey experiment with a fixed delay between the two $\pi/2$ pulses. The duration of the pseudo-readout pulse was fixed at $\tau=210$ ns to eliminate the beating pattern observed in Fig.~\ref{fig:ramsey}. The corresponding pulse sequence is shown in Fig.~\ref{fig:ProbeInducedDephasing:a}. The phase $\theta$ of the second $\pi/2$ pulse was swept from $0$ to $2\pi$, producing Ramsey oscillations of the form $c \sin(\theta+\theta_0)$. The oscillation contrast $c$ was used to extract the excited-state population $P_e = 2c$, while the phase offset $\theta_0$ is due to qubit frequency shifts. The amplitude and detuning of the pseudo-readout pulse were varied. As shown in Figs.~\ref{fig:ProbeInducedDephasing:d} and \ref{fig:ProbeInducedDephasing:f}, implementing DRAG substantially suppresses the additional errors induced by stray readout pulses.

\begin{figure*}[t]
    \centering
    \includegraphics[width=0.98\linewidth]{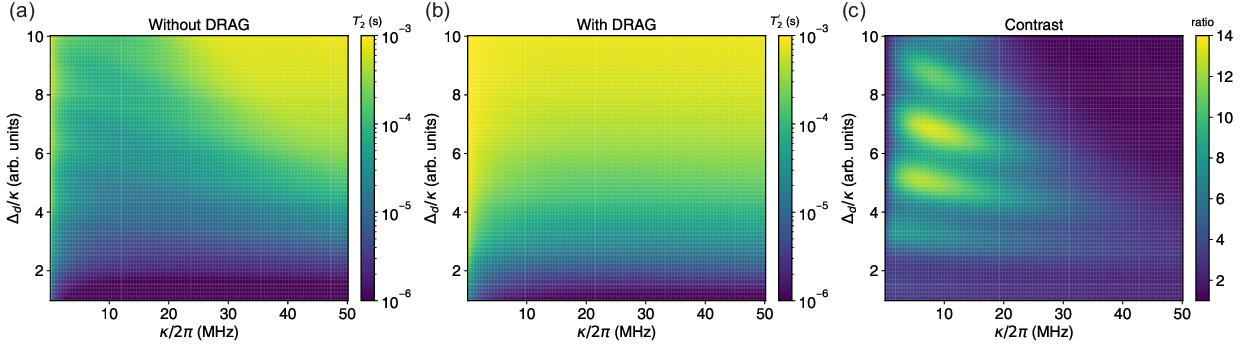}
    \subfigure{\label{fig:fig4:a}}
    \subfigure{\label{fig:fig4:b}}
    \subfigure{\label{fig:fig4:c}}
    \caption{
    \textbf{Dependence of readout-induced $T'_2$ on normalized probe detuning $\Delta_d/\kappa$ and resonator linewidth $\kappa$. }
    Theoretical evaluation of readout-induced transverse relaxation time obtained by scanning the pseudo-readout detuning and resonator linewidth using Eq.~\ref{eq:dephasingratenew}. (a) Without DRAG. (b) With DRAG. (c) Ratio of (b) to (a), illustrating the enhancement in $T'_2$ achieved by DRAG. }
    \label{fig:fig4}
\end{figure*}

The theoretical calculations exhibit excellent agreement with the experimental data as shown in Figs.~\ref{fig:ProbeInducedDephasing:c}-\ref{fig:ProbeInducedDephasing:f}, both with and without DRAG applied. 
To develop the theoretical model, we employ a simplified expression that relates the excited-state population $P_e$ to the total transverse relaxation rate $\Gamma'_{2}$ as $P_e=\exp(-\tau\Gamma'_{2})$, where $\Gamma'_{2}=\Gamma_{2}+\Delta\Gamma_\varphi$. Here, $\Gamma_{2}=1/T_2$ is the intrinsic transverse relaxation rate in the absence of the pseudo-readout pulse and $\Delta\Gamma_\varphi$ is the additional dephasing rate induced by the pseudo-readout drive.
We then consider the case where the resonator is driven by a monochromatic field of amplitude $A$. Due to the dispersive qubit-resonator interaction, described by the Hamiltonian term $\chi \hat{a}^\dagger \hat{a}\sigma_z$ with $\chi$ denoting the dispersive shift, the qubit dephasing rate induced by intracavity photons is given by \cite{Gambetta2008Quantum,bultink2018general,blais2021circuit}:
\begin{equation}\label{eq:dephasingrate}
\Delta\Gamma_{\varphi} = \frac{2 |\varepsilon|^2 \chi^2 \kappa}{\left[(\Delta_d+\chi)^2+(\kappa / 2)^2\right]\left[(\Delta_d-\chi)^2+(\kappa / 2)^2\right]},
\end{equation}
where $\varepsilon = i\sqrt{\kappa}A$, and $\Delta_d = \omega_r - \omega_d$ denotes the detuning between the resonator frequency $\omega_r$ and the drive frequency $\omega_d$. For simplicity, we use the steady-state expression valid in the long-time limit, since the measurement time $t_m$ in our experiment exceeds the photon decay time ($t_m \sim 3/\kappa$).

Then, we account for the effect of the readout pulse’s spectral width. The Fourier transform of the readout pulse envelope yields the spectrum $\varepsilon(\Delta - \Delta_d)$, where $\Delta=\omega_r-\omega$ denotes the detuning between the resonator frequency $\omega_r$ and each spectral component $\omega$ of the readout pulse. Incorporating spectral broadening, the constant drive strength $\varepsilon$ in Eq.~\ref{eq:dephasingrate} is replaced by the frequency-dependent strength $\varepsilon(\Delta - \Delta_d)$, and the resulting dephasing rate is therefore expressed as
\begin{equation}\label{eq:dephasingratenew}
\Delta\Gamma'_{\varphi} = \int \frac{2 |\varepsilon(\Delta - \Delta_d)|^2 \chi^2 \kappa}{\left[(\Delta_d+\chi)^2+(\kappa / 2)^2\right]\left[(\Delta_d-\chi)^2+(\kappa / 2)^2\right]}d\Delta.
\end{equation}

In our theoretical results, the only fitting parameter is the proportionality coefficient between $\varepsilon$ and the normalized probe amplitude. This coefficient is kept fixed for both the DRAG and non-DRAG cases. In addition, the $S_{21}$ response of the readout line was measured separately and fitted to extract $2\chi$ and $\kappa$, which were then used in the theoretical calculations [Figs.~\ref{fig:ProbeInducedDephasing:e} and \ref{fig:ProbeInducedDephasing:f}].

In the DRAG-applied data in Figs.~\ref{fig:ProbeInducedDephasing:d} and \ref{fig:ProbeInducedDephasing:f}, the data at zero detuning is omitted. This is because the DRAG technique is not defined at zero detuning, and no readout pulse was applied in this case. In practice, frequency spacing in multiplexed readout is never exactly zero and is typically larger than the resonator linewidth to avoid spectral overlap of resonators. The frequency spacing is typically tens to hundreds of megahertz \cite{heinsoo2018rapid,spring2025fast}, and the DRAG method remains applicable under realistic operating conditions. We also note that two peaks with elevated dephasing error appear near $\pm$40 MHz at large probe amplitudes, the origin of which remains unclear and warrants further investigation.

To study the effect of the resonator linewidth---which is now typically on the order of 10 MHz in fast-readout systems---we calculated the readout-induced $T'_{2}=1/\Gamma'_{2}$ based on Eq. \ref{eq:dephasingratenew}. In the calculation, we set $2\chi = \kappa$ and use a readout pulse with a plateau of $3/\kappa$, 5-ns rising and falling edges, and a fixed driving strength. The intrinsic qubit coherence time is taken to be $T_2=1$ ms in the absence of the readout pulse. The results in Fig.~\ref{fig:fig4} show that, under our simulation parameters, the application of DRAG yields an enhancement factor in the transverse relaxation time $T_2'$ ranging from 1 to 13.2. The maximum enhancement is achieved at a normalized detuning of 6.8$\kappa$ and a resonator linewidth of 7.9 MHz, as shown in Fig.~\ref{fig:fig4:c}. Accordingly, for commonly used resonator linewidths (on the order of 10 MHz), applying DRAG reduces the required frequency spacing and thereby improves the achievable readout multiplexing within a given bandwidth.

In summary, we have shown that the spectral broadening of short readout pulses introduces additional crosstalk errors, which can be strongly suppressed using the DRAG technique. Our approach remains hardware-friendly and offers a practical route toward scalable multiplexed readout. Looking forward, several extensions are promising. First, probe-pulse DRAG could be combined with individual Purcell-filter architectures, where spectral overlap between readout resonators is already suppressed and residual crosstalk is dominated by the broad spectral width of short pulses, to relax frequency-spacing requirements. Second, in systems with both positively and negatively detuned neighboring resonators, the use of the dual DRAG method\cite{li2025Universal,wang2025suppressing} allows simultaneous suppression of spectral components on both sides of the main resonance. Furthermore, investigating the impact of advanced pulse-shaping techniques, such as the DRAG method, on the quantum non-demolition (QND) nature of the measurement is an important avenue for future work. Finally, integration with other pulse-shaping strategies---such as ring-up and active ring-down techniques \cite{mcclure2016rapid,jerger2024dispersive,hazra2025benchmarking,chatterjee2025enhanced}---is a promising pathway for further improving readout fidelity and speed.

\begin{acknowledgments}
This work was supported by the Beijing Natural
Science Foundation (Grants No. 4262019 and No. JQ25014), the National Natural Science Foundation of China (Grants No. 12204228, No. 12322413, No. 92476206, and No. 92365206), and the Innovation Program for Quantum Science and Technology (Grant No. 2021ZD0301802)
\end{acknowledgments}

\bibliography{aipsamp}

\end{document}